# Ethics Readiness of Technology:
# The case for aligning ethical approaches with technological maturity

Eline de Jong[1*]

**Abstract**

The ethics of emerging technologies faces an anticipation dilemma: engaging too early risks overly speculative concerns, while engaging too late may forfeit the chance to shape a technology's trajectory. Despite various methods to address this challenge, no framework exists to assess their suitability across different stages of technological development. This paper proposes such a framework. I conceptualise two main ethical approaches: *outcomes-oriented ethics*, which assesses the potential consequences of a technology's materialisation, and *meaning-oriented ethics*, which examines how (social) meaning is attributed to a technology. **I argue that the strengths and limitations of outcomes- and meaning-oriented ethics depend on the uncertainties surrounding a technology, which shift as it matures.** To capture this evolution, I introduce the concept of *ethics readiness*—the readiness of a technology to undergo detailed ethical scrutiny. Building on the widely known Technology Readiness Levels (TRLs), I propose Ethics Readiness Levels (ERLs) to illustrate how the suitability of ethical approaches evolves with a technology's development. At lower ERLs, where uncertainties are most pronounced, meaning-oriented ethics proves more effective; while at higher ERLs, as impacts become clearer, outcomes-oriented ethics gains relevance. By linking Ethics Readiness to Technology Readiness, this framework underscores that the appropriateness of ethical approaches evolves alongside technological maturity, ensuring scrutiny remains grounded and relevant. Finally, I demonstrate the practical value of this framework by applying it to quantum technologies, showing how Ethics Readiness can guide effective ethical engagement.

**Acknowledgements:** I am grateful to Dr Clare Shelley-Egan and Dr Sebastian De Haro for their constructive feedback on earlier drafts of this article. I also thank the organisers and participants of various workshops and conferences—particularly the KIT-ITAS Responsible Quantum Technology Conference 2025—for the opportunity to present and discuss this work. The insights and discussions from these exchanges significantly contributed to the refinement of the paper.

**Funding**: This publication is an outcome of the project Quantum Impact on Societal Security, project number NWA.1436.20.002, which is funded by the Dutch Research Council (NWO), The quantum/nanorevolution.

**Conflict of interest**: All authors declare that they have no conflicts of interest.

[1,2] Institute for Logic, Language and Computation; Institute of Physics; Qusoft Research Center for Quantum Software; University of Amsterdam, The Netherlands

*Corresponding author: e.l.dejong@uva.nl

## 1. Introduction: Addressing the dilemma between 'too early' and 'too late'

The idea that ethics should not be an afterthought is taken by many to be a guiding principle in anticipating ethical issues raised by emerging technologies. The underlying idea is that ethical and broader societal concerns, including the desirability of the technology in question, can best be addressed as early as possible, before accidents happen or undesirable trends become entrenched, and while the technology is still malleable. Thus, proactive ethical engagement is essential to prevent ethics from arriving 'too late' (Grunwald, 2005: p.200)—when negative consequences may already be manifest, or shaping the design and deployment of a technology becomes significantly harder.

Quantum technologies,[2] a family of emerging technologies, offer a prime opportunity to enact this proactive stance. Given their potential societal impact, calls have been made for a "responsible approach," emphasising early identification and mitigation of ethical issues (Coenen & Grunwald, 2017; Inglesant et al., 2021; Ten Holter et al., 2023; Seskir et al., 2023; Kop et al. 2023; Kop et al. 2024; Gasser, De Jong & Kop, 2024). With lessons learned from fields like nanotechnology and AI, the ambition to "do things right from the outset" has intensified and shapes the emerging debate about the ethics of quantum technologies.

While the early stage of quantum technologies offers the chance to effectively shape their trajectory, it also presents a knowledge problem. There is still considerable uncertainty about how and when and even if promising technologies like quantum computing and quantum communication will materialise. Given this uncertainty and the inherent complexity and unpredictability of technological innovation generally, the anticipation imperative may force early-stage ethical discussion to rely on a range of assumptions. Facing the risk of becoming overly speculative, ethics thus may also come in 'too early'. How, then, can we balance the timing of ethical considerations for emerging technologies like quantum?

Over the last few decades, a wide range of methods has been developed for the ethical study of early-stage technologies.[3] However, there has been ongoing debate about the strengths and weaknesses of these methods, with none emerging as dominant (Reijers et al., 2018). An overarching framework for evaluating the applicability or *appropriateness* of these methods—understood here as their suitability given contextual constraints and goals—remains lacking. This paper seeks to fill this gap by providing such a framework, one that links the appropriateness of ethical methods to a technology's stage of development. In particular, it addresses the key question: How can we align our approach to ethics with the stage of technological development to ensure timely and meaningful ethical consideration? In other words, how can we ensure that ethical considerations in emerging technology are neither premature nor overdue?

To answer that question, I conceptualise two main ethical approaches: *outcomes-oriented ethics*, which assesses the potential consequences of a technology's materialisation, and *meaning-oriented ethics*, which examines how (social) meaning is attributed to a technology (Section 2). I then take this conceptualisation further to argue that the appropriateness of these approaches depends on the uncertainties surrounding a technology, which shift as it matures. To capture this evolution, I introduce the concept of Ethics Readiness—the readiness of a technology to undergo detailed ethical scrutiny (Section 3). Building on the widely known Technology Readiness Levels (TRLs), I propose Ethics Readiness Levels (ERLs), which indicate the types of ethical questions that are appropriate at different stages of a technology's development. The resulting framework helps assess the suitability of ethical approaches based on a technology's maturity and can guide ethics in practice. Finally, to explore the practical implications of the framework, I apply it tentatively to the case of quantum technologies (Section 4).

---

[2] This paper focuses exclusively on second-generation quantum technologies, which harness quantum mechanical phenomena for emerging applications such as enhanced sensing, advanced simulation and computation, and novel communication systems. By contrast, first-generation quantum technologies include well-established innovations like transistors, lasers, and magnetic resonance imaging (MRI).
[3] Reijers et al. (2018) identify 35 different methods for practicing ethics in research and innovation.

## 2. Outcomes- and meaning-oriented approaches to the ethics of emerging technology

In this section, I explore the landscape of methods in the ethics of emerging technologies. I begin by identifying the central challenge in this field: addressing the epistemological issue of uncertainty and ignorance (2.1). Next, from the literature I distil two primary approaches to the ethics of emerging technologies: those focused on anticipating future outcomes (2.2) and those centred on understanding the present meanings attributed to technology (2.3). Building on this distinction, I argue that the appropriateness of these approaches varies depending on the specific stage of technological development (2.4). This analysis lays the groundwork for the concept of 'Ethics Readiness,' which is further developed in Section 3.

### 2.1. Upstream-ethics and the problem of uncertainty

The ethical study of emerging technologies and their impacts—commonly referred to as the ethics of new and emerging science and technology (NEST) or simply NEST-ethics (Swierstra & Rip, 2007)—is inherently anticipatory. To understand this anticipatory character, it can be helpful to imagine technological innovation as a 'stream': to avert or mitigate undesirable outcomes and promote desirable ones 'downstream', it is essential to make informed decisions 'upstream', early in the innovation process. This 'upstream-ethics', as I will refer to the ethics of early-stage technologies, aligns with the broader paradigm of responsible research and innovation (RRI), which is generally geared towards anticipation and integrating societal considerations throughout the research and innovation process (Ryan & Blok, 2023; Genus & Stirling, 2018; Wickson and Carew, 2014).

Anticipating ethical concerns highlights a fundamental challenge to upstream ethics and RRI more generally: navigating an unknown future (Lucivero, Swierstra, & Boenink, 2011; Sollie 2007). As Swierstra and Rip (2007: p.6) aptly state: "Working with a novelty necessarily means venturing into the unknown." This uncertainty hallmarks all NEST-ethics, ranging from debates about Artificial Intelligence and quantum computers, to lab-grown meat, cloning, and genetic engineering. The complexity and unpredictability of technological innovation implies uncertainty about a technology's trajectory and ignorance about its consequences (Ferrari, 2010). In this sense, anticipating ethical concerns is '*doubly fictional*' (Rip & te Kulve, 2008): grappling with yet-to-materialise technologies and their yet-to-manifest implications.

This epistemological problem is rooted in the very character of innovation itself, which is inherently geared toward actively seeking novelty (Steen, Sand & Van de Poel, 2021; Nordmann, 2014). This aligns with Lucivero, Swierstra and Boenink (2011), who state that the object of ethical assessment of emerging technology is, by definition, "elusive" (p.129). While uncertainty and unpredictability are intrinsic to innovation, this elusiveness is most pronounced during the earliest stages of research and development.

The counterpart of this early-stage knowledge problem is a late-stage power problem: as the technology matures and becomes embedded in society, the capacity to effectively adapt its design and deployment diminish. This tension is captured by Collingridge's (1980) famous dilemma of social control of technology, which highlights the trade-off between early-stage uncertainty and late-stage entrenchment. This dilemma has an ethical variant, with ethics either coming in 'too soon', "being too speculative to be reliable", or coming in 'too late' to effectively shape the technology's impacts (see also Kudina & Verbeek, 2019).

One often-advocated response to the dilemma of social and ethical control focuses on fostering 'technological flexibility' or corrigibility (Collingridge, 1980; Joly, 2015; Genus & Stirling, 2018). This involves designing technology to be adaptable and responsive to feedback as new social or ethical concerns arise, theoretically allowing society to retain a measure of control. Yet realising such flexibility, and ensuring it adequately addresses ethical concerns, presents challenges. First, the design of flexible technologies itself requires foresight about potential ethical issues, which revisits the knowledge problem. Second, flexibility often reduces to periodic evaluation rather than meaningful adaptability, which is in general not sufficient. Some ethical issues, particularly those warranting precautionary action, cannot be fully mitigated by flexible design. For instance, the cybersecurity risks that large-scale quantum computing might bring cannot be fully mitigated through adaptability alone; the technology's core capabilities could undermine current security protocols. Technological flexibility is thus not a panacea, and it risks backgrounding the anticipatory ethics necessary for emerging technologies.

Anticipating a late-stage power problem—where the opportunity to shape a technology's trajectory is lost due to accumulated, practically irreversible decisions—inherently involves addressing the early-stage knowledge problem and thus finding ways to responsibly navigate uncertainty. Over the past two decades, numerous methods have been proposed to tackle this challenge, with a notable increase in the last ten years (Reijers et al., 2018). These methods—commonly referred to as "ex ante methods"—focus on engaging with ethical issues before technologies have materialised, at the earliest stages of research and innovation (Reijers et al., 2018; Urueña, 2024). In the following sections, I distinguish between two primary approaches within this upstream ethics, each offering distinct strategies for addressing the uncertainties inherent in early-stage anticipation.

*2.2. Outcomes-oriented ethics: Engaging with possible futures*

One strategy to engage with unknown futures is to envision *possible futures* and take these as the object of ethical study and debate. Prominent examples include anticipatory technology ethics (Brey, 2012), ETICA (EThical Issues of emerging iCt Applications) (Stahl & Flick, 2011), ethical technology assessment (Palm & Hansson, 2006), ethical-constructive technology assessment (Kiran, Verbeek, & Oudshoorn, 2015), and techno-moral scenarios (Boenink, Swierstra, & Stemerding, 2010). Methods of this sort all employ a type of ethical foresight analysis (Floridi & Strait, 2020): they aim to envision a future technology and the possible outcomes it might instigate. These possible futures are then taken as the object of ethical scrutiny.

Yet, as multiple scholars have noted, early ethical reflection can become speculative when uncertainty about a technology's development and applications remains high (Nordmann, 2007; Swierstra & Rip, 2007; Nordmann & Rip, 2009; Brey, 2012; Ferrari, 2010; Grunwald, 2017; Reijers et al., 2018). This critique of speculative ethics has most prominently been voiced by Nordmann (2007; 2014). Central to Nordmann's critique is that ethics excessively leaps ahead of science. By relying on too many assumptions, ethics 'foreshortens the conditional', treating hypothetical scenarios as presenting actual ethical issues (Nordmann, 2007: p.32). This "thinking ahead too much" (Gilbert & Goddard, 2014) risks an arbitrary focus, potentially diverting intellectual resources to unlikely scenarios while neglecting pressing ethical issues.

The critique of speculative ethics raises questions about framing the potential ramifications of emerging technologies as 'risks'—commonly defined as a probability multiplied by impact—when the probabilities involved are sheer unknowns. Efforts to mitigate speculative risks might divert attention from other, more pressing, issues. However, anticipating risks is fundamental to ethics' role in shaping technology, thus requiring it to leap ahead of scientific and technological developments. Importantly, when the expected impact of a technology is potentially vast, even low-probability scenarios might warrant serious ethical consideration (Price, 2024). We will return to this issue in Section 4.

Acknowledging the need to think ahead, scholars like Roache (2008) and Urueña (2022; 2023) argue that ethical debate should not be confined to current technology: they defend the importance and possibility of anticipating possible future scenarios. Likewise, Grunwald (2010) and Selin (2014) hold that engaging with possible futures is a useful reflexive heuristics. Even critics of speculative ethics admit that controlled or responsible speculation is essential to ethical enquiry (Nordmann & Rip, 2009). Thus, upstream ethics must engage with possible futures—yet the challenge remains how to do so effectively, while carefully avoiding the pitfalls of speculation.

To address this challenge, it is crucial to examine the key target of critiques of speculative ethics. These critiques often focus on the predictive mode of anticipation, which centres on future *outcomes*. NEST-ethics has indeed been associated with consequentialist patterns of reasoning that evaluate actions based on their results rather than the actions themselves or the intentions behind them (Swierstra & Rip, 2007; Ferrari, 2013; Grunwald, 2020): "The new and emerging technology is deemed desirable, or not, because its consequences are desirable, or not." (Swierstra & Rip, 2007: p.11) However, while this consequences-focused approach aligns with consequentialist reasoning, it does not necessarily imply strict consequentialism, which locates all normative value solely in outcomes. For this reason, I wish to refer to the consequentialist tendency in upstream ethics more broadly as 'outcomes oriented.'

The outcomes-oriented approach to ethics, or in short 'outcomes-oriented ethics', typically focuses on the consequences of a technology, which can occur at multiple levels. Taking Artificial Intelligence (AI) as an example, this approach would focus ethical enquiry on the implications of design choices, and on potential uses, as well as the direct effects of that use. Additionally, this focus on outcomes could expand to more indirect societal impacts, such as changes in social dynamics, deskilling, and privacy concerns.

Several authors argue that ethics focused on consequences or outcomes faces challenges due to the inherent unpredictability of innovation (Grinbaum & Groves, 2013; Nordmann, 2007; Sand, 2018; Steen, Sand & Van de Poel, 2021; Vallor, 2016). An emphasis on potential outcomes assumes a level of foresight that clashes with the uncertainty of early technological development. NEST-ethics thus risks excessive speculation when assessing possible outcomes at a nascent stage. In essence, the critique of speculative ethics targets attempts to conduct 'downstream ethics'—focused on assessing specific consequences of a concrete technology—when still at a very 'upstream' point in its development.

Beyond predictive anticipation, there are alternative modes of engaging with technological futures. In responsible innovation literature, authors like Roache (2008) and Urueña (2022) highlight the value of explorative, strategic, and critical-hermeneutic modes of anticipation. Shifting from predictive to hermeneutic anticipation at the earliest stages of innovation suggests moving away from a focus on speculative outcomes and toward examining how, by whom, and under what assumptions these outcomes are envisioned and given significance. This alternative perspective on anticipating and engaging with possible futures introduces a second approach to ethics, which the next section examines in more detail.

*2.3. Meaning-oriented ethics: Assigning meaning to future technologies*

Besides the outcomes-oriented approach, a second approach has emerged within the ethics of emerging technology. This approach seeks to mitigate 'speculative excess' by grounding ethical enquiry more firmly in the present. Some of them aim to reduce speculation by basing ethical enquiry in concrete practices. Examples are real-time technology assessment (Guston & Sarewitz, 2020), and experimental ethics (van de Poel, 2016). Yet, because these methods depend on a technology's actual level of development, they may face limitations in early-stage anticipation.

Other methods that seek to ground ethics in the present focus on "the future as it exists already" (Nordmann & Grunwald, 2023). Since emerging technologies are *emerging*, they predominantly exist in the form of visions, promises, and expectations (Borup et al., 2006; Lucivero, Swierstra, & Boenink, 2011). These "future-oriented abstractions" (Borup et al. 2006) capture—individual or collective—ideas about how the future might look like, "often expressed in the semantic of intentions, goals, hopes or proposals." (Lucivero, Swierstra, & Boenink, 2011: p.131). Also, these abstractions tend to focus on the generic qualities of a technology, rather than specific artefacts and applications.

According to Grunwald (2017), envisioning 'technology futures'—imaginaries of what society might look like when a specific technology materialises—entails assigning *meaning* to a technology; the technology becomes something to appraise. Indeed, expectations and visions do inform us about the perceived desirability and acceptability of technologies. These assignments of meaning are value-laden, normative acts and hence, can be subjected to ethical reflection. In fact, Grunwald argues that it is crucial to do so, because meaning-assignment activities—the way we think and talk about the future—set the stage and continue to influence the subsequent debate (Grunwald, 2023; see also Van der Burg, 2014). Or as Lucivero, Swierstra and Boenink (2011) put it: "[T]here is a self-referential loop between the present and the future: the way in which we *describe* the future will determine *how* the future will be." If we seek to study an emerging technology, we thus need to study how it is being perceived and ethically assess how it comes to matter.

Several methods have been proposed for the ethical study of meaning-assignment by critical reflection on "sociotechnical imaginaries" (Jasanoff, 2020). These include vision assessment (Schneider et al., 2023, Nordmann, 2007; 2014; Grunwald, 2004; 2014; 2020), reflection on metaphysical-research programmes (Ferrari, 2010), and hermeneutic technology assessment (Nordmann & Grunwald, 2023). What connects these methods, is the idea that critical examination of sociotechnical imaginaries can help identifying the values, assumptions, and power dynamics that shape technological innovation (Urueña, 2022). It also brings into view *who* shapes these visions and *for whom*: in other words, critical examination of visions can reveal the 'impact-makers,' those with the power to shape visions and a technology's trajectory, and what we may call 'impact-takers', those who lack that power but are likely to experience the impact of a technology.

I will collectively refer to the ethical methods focused on meaning-assignment and fundamental questions about the sociotechnical imaginaries surrounding a technology as the 'meaning-oriented' approach to ethics or, in short, 'meaning-oriented ethics.' Meaning-oriented ethics aims to facilitate ethical discussion at the early stages of a technological project, while reducing speculation about how the future will look like. The analysis of meaning-assignment is not hindered by the problem of uncertainty in the same way outcomes-oriented ethics is. Although 'techno-visionary futures' (Grunwald, 2023), or in short 'techno-visions', involve images of what the future with a specific technology may look like, it is not the results themselves that are the object of ethical scrutiny, but rather the act of envisaging them. Nonetheless, ethical study of visions, promises, and expectations necessarily includes the assessment of the quality of the claims that constitute them—not only taking technological feasibility into account, but also the perceived social usability and desirability (Lucivero, Swierstra, & Boenink, 2011). In that way, the probability and plausibility of technology futures become the object of ethical scrutiny, and responsibly dealing with uncertainty takes shape as fostering 'responsible representation' (Nordmann & Rip, 2009).

Besides this hermeneutic approach, which focuses on the process of envisioning, meaning-assignment can also be studied from a conceptual point of view. Lunshof and Rijssenbeek (2024) argue that the earliest phases of technology development should be considered 'pre-ethical'. At this stage of fundamental research and early development, the technology is considered too premature to be subjected to genuine ethical analysis. To prepare for such ethical analysis at later stages, it is crucial to critically reflect on the meaning of key concepts that might be disrupted by the future technology. Instead of labelling such conceptual analysis as pre-ethical, I assert that this kind of research can be subsumed under meaning-oriented ethics since it focuses on how we attribute meaning to a technology through (revised or 're-engineered') concepts.

Revisiting the example of AI, meaning-oriented ethics would focus on examining the values that underpin its development, questioning the desirability of the envisioned futures, and critically reflecting on how the deployment of AI may affect existing power structures and vice versa. It could also involve exploring how AI challenges established definitions of values and disrupts fundamental concepts, like autonomy, control, and humanity itself. This type of ethical analysis thus focuses more on structural issues associated with a future technology, instead of the implications of specific applications.

Figure 1 gives a non-exhaustive overview of the outcomes- and meaning-oriented approaches to ethics.

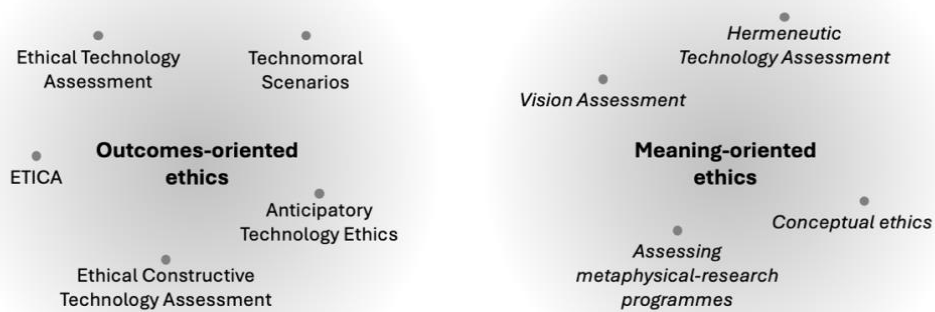

*Figure 1. Distinction between outcomes- and meaning-oriented approaches to the ethics of emerging technologies, and some of their paradigmatic methods.*

*2.4. Evaluating ethical approaches to emerging technology: Introducing a temporal dimension*

The outcomes- and meaning oriented approaches to ethics deal with uncertainty in different ways. Outcomes-oriented ethics tackles uncertainty by speculating about potential impacts, creating an object of anticipation. However, this can lead to speculative overreach, especially in the early stages. In contrast, meaning-oriented ethics focuses on grounding ethical analysis in the present, examining the visions, promises, and assumptions that shape how technologies are imagined. This focus brings the analysis closer to the social and cultural contexts in which these imaginaries are formed, offering a more immediate connection to the present. However, as technologies advance, assessing potential impacts becomes crucial, highlighting the limitations of meaning-oriented ethics.

The strengths and limitations of both approaches are closely tied to *timing*. As technologies mature, the epistemic conditions change, altering their suitability and relevance. Thus, their appropriateness or meaningful applicability is highly dependent on the specific stage of technological development. Urueña (2024) touches on a similar idea by linking different "modes of anticipation" to variables such as the timing of anticipation practices (ex-ante, ex-dure, ex-post). However, the concept that ethical approaches should adapt alongside technological development remains underexplored. In discussions of upstream or ex-ante ethics (Brey, 2012; Floridi & Strait, 2020; Reijers et al., 2018), the term "emerging" often lumps together stages of research and development that differ in their epistemic conditions and thus their ethical significance. Thus the ethical study of emerging technologies would benefit from incorporating this temporal dimension into the frameworks for assessing the appropriateness of different approaches.

Similarly, it is crucial to integrate the connection between a technology's stage of development and the appropriate level of analysis. While useful distinctions have been made—for example, Brey's (2012) focus on technology, artefacts, and applications, and Ryan et al.'s (2024) differentiation between the microlevel (artefact), mesolevel (socio-political structures), and macrolevel (ontological considerations)— the relevance of each level in relation to the stage of technological development remains implicit. However, it is reasonable to assume that the stage of development significantly impacts the relevance of each level. For example, analysing artefacts and applications typically assumes a more advanced stage of development, while earlier stages may require broader considerations, such as those at the technological level.

The framework I propose in the next section introduces a temporal dimension into the evaluation of ethical approaches and the appropriate level of analysis. I argue that different stages of technological development allow for different kinds and levels of ethical discussion. Using the Technology Readiness Level (TRL) scale, I link the relevance of outcomes- and meaning-oriented ethics to specific levels of maturation, aligning ethical scrutiny with the actual stage of technological development. Incorporating this temporal dimension into the evaluation of ethical approaches aligns with Collingridge's (1980) incrementalist approach to the dilemma of control, which stresses that the assessment of technology should evolve alongside its development.

## 3. The ethics readiness of technology and its implications for ethical study

This section develops a framework that explicitly and systematically links the appropriateness of outcomes- and meaning-oriented ethics to a technology's stage of development. It begins by discussing the Technology Readiness Level (TRL) scale and its societal and ethical dimension (3.1). Drawing on this scale and the discussions in Section 2, I introduce the concept of "ethics readiness" as a measure of a technology's preparedness for detailed ethical scrutiny (3.2). Next, I link ethics readiness with the TRL scale, resulting in the Ethics Readiness Level (ERL) scale (3.3). This scale highlights how the appropriateness of ethical approaches depends on the technology's maturity.

### 3.1. Technology Readiness and Societal Readiness

If the specific stage of a technology's development is decisive for determining the appropriate ethical approach, identifying this stage becomes crucial. The Technology Readiness Level (TRL) scale provides a concrete framework for distinguishing between these stages. Originally developed by NASA, this framework assesses the maturity level of a specific technology on a scale that defines 'readiness' in relation to market viability. The central idea is that typically, new technologies go through various stages of the TRL scale in their life cycle.

The TRL framework is widely adopted as a tool for decision-making, for example by the European Commission (EC) in the Horizon Europe funding programmes. In the EC's version, the TRL scale comprises nine levels of maturation, which can be categorised into four phases (APRE & CDTI, n.d.; Netherlands Enterprise Agency, 2022):

- **Discovery Phase (TRL 1-3)**: This exploratory phase includes fundamental research, applied research, and the creation of a proof-of-concept. By TRL 3, the technology is considered scientifically feasible.
- **Development Phase (TRL 4-6)**: Encompassing testing, validation, and prototype demonstration, this phase ensures the technology is feasible and reliable from a technological standpoint by TRL 6.
- **Demonstration Phase (TRL 7-8)**: This phase involves demonstrating the technology in its operational environment and finalising it for implementation. At TRL 8, the technology is both technically and commercially ready.
- **Deployment Phase (TRL 9)**: The final phase, where the technology is scaled and introduced to the market, marking its readiness as a commercial product or service.

The TRL scale predominantly focuses on readiness from a technological perspective, assessing whether *the technology* is ready for deployment in society ('technological feasibility'). The framework does not explicitly and systematically account for whether *society* is ready for the uptake of the technology at hand. However, to effectively and responsibly embed a technology in society, it is crucial to consider society's 'readiness' as well, including regulatory frameworks, broader sociotechnical ecosystems, and attitudes towards the technology (WRR, 2019; De Jong, 2022). Neglecting society's readiness may obstruct successful uptake and lead to resistance, governance problems, and broader issues of nonalignment with public values.

To gauge societal readiness, Bernstein et al. (2022) developed the 'Societal Readiness Thinking Tool'. The tool is based on the idea that societal readiness is an iterative process of addressing societal concerns during the technology life cycle. These societal concerns are categorised in five dimensions: public engagement, open access, science education, gender, ethics, and governance. By presenting sets of guiding questions for different phases of the TRL scale, the tool seeks to integrate thinking about societal concerns at different stages of the innovation process.

Similarly, Bruno et al. (2020) have proposed an extension of the TRL scale in three directions, resulting in scales for Legal, Organisational and Societal Readiness Levels. Just like technological innovation itself,

Bruno et al. argue that compatibility with legal frameworks, organisational structures and societal needs also progresses through various stages of maturity. In their framework, the levels of legal, organisational and societal readiness mirror the technology readiness levels, resulting in nine stages. These stages are primarily explained in terms of 'readiness to adopt' a particular technology, from a legal, organisational and societal point of view. In other words, the focus of this framework is on identifying legal, organisational, and societal (or social) 'the enablers and barriers' for the take-up of a particular technology.

Both frameworks can be seen as efforts to expand and contextualise the original Technology Readiness Level (TRL) scale. They broaden the range of considerations included in assessing a technology's readiness for deployment in society. Notably, both frameworks incorporate ethical dimensions: Bernstein et al. (2022) identify ethics as one of the 'dimensions' of societal readiness, while Bruno et al. (2020) include ethical considerations within the scope of legal compliance. These contributions support the idea that ethical discussions evolve alongside progress on the TRL scale.

However, the idea that different stages of technology development allow for—and even require—different approaches to ethics, remains implicit in these frameworks. Building on the previous discussion about the suitability of specific ethical approaches at various phases of technological development, there is a strong case for explicitly linking the type or 'mode' of ethical considerations to the TRL scale. Beyond its current function of guiding innovation policies and serving as a decision-making tool, the TRL scale could also help to determine the appropriate ethical approach for emerging technologies, such as quantum technology.

*3.2. Ethics Readiness*

The proposals by Bernstein et al. (2022) and Bruno et al. (2020) to complement the TRL scale with a societal readiness maturity scale lay the groundwork for viewing NEST ethics as evolving alongside technological development. Building on this idea, I introduce the concept of *ethics readiness*[4].

I define "ethics readiness" as the extent to which a technology is prepared to undergo detailed ethical scrutiny. While ethical considerations are relevant from the outset of technological innovation, it is important to distinguish between the general need for ethics and the appropriateness of specific ethical approaches at different stages of technological development. Ethics cannot come "too early" in a general sense, but certain types of (detailed) ethical questions may be premature at specific stages.

The ethics readiness of a technology evolves along its technological development. The ethical analysis it supports thus becomes increasingly detailed and specific over time. In the early phases, particularly during the conceptual stage, ethics readiness is typically low due to high uncertainties and a lack of concrete information. As a result, the technology is not yet ready for detailed ethical scrutiny. However, this early stage allows for more abstract ethical analysis, such as that associated with meaning-oriented ethics. As we have seen in Section 2.3, meaning-oriented ethics examines the broader visions, promises, and assumptions shaping a technology's trajectory. While this approach rigorously addresses systemic and abstract concerns, it is less granular in addressing specific features of the technology itself.

In contrast, outcomes-oriented ethics focuses on specific features, applications, and impacts of the technology. This approach inherently involves more detailed ethical scrutiny, as it examines concrete aspects of the technology and their potential consequences (see Section 2.2). As the technology matures and uncertainties decrease, its ethics readiness increases, enabling more detailed and specific analyses typically associated with outcomes-oriented ethics.

---

[4] I use "ethics readiness" instead of "ethical readiness" because the latter may suggest that a technology is ethically ready in the sense of aligned with ethical values. Ethics readiness, on the other hand, encompasses the readiness of a technology *for* ethics.

The idea of a technology's evolving ethics readiness can be further understood through the lens of "critical-reflective affordances". In a recent paper, Urueña (2024) introduces this term to describe the opportunities that certain modes of anticipation provide for critical reflection on future technologies. These affordances determine the kinds of ethical questions that different frameworks of anticipation allow us to ask, where some modes are better suited to certain stages of technological development than others.

Applied to technology[5], critical-reflective affordances describe the *ethical questions and discussions* afforded by a technology at a given stage of development. Each stage of a technology's development offers different entry points for ethical reflection. These entry points are the affordances for critical reflection: the possibilities for scrutiny shaped by the technology's state of development. As technologies mature, these affordances evolve, driven by increasing availability of concrete information and a narrowing scope of uncertainties. At early stages, when a technology is still an idea, it affords reflection on its potential purpose, societal implications, and alignment with values. As it evolves into a working prototype, it affords consideration of design choices, applications, and potential impacts on users or society. This understanding adds depth to the concept of ethics readiness: it reflects the type and specificity of ethical questions a technology affords at specific stages of its development.

### 3.3. Pairing Technology Readiness with Ethics Readiness

To assess a technology's ethics readiness, I propose the Ethics Readiness Level (ERL) scale, aligning with the Technology Readiness Level (TRL) scale. This means that a low TRL typically corresponds to a low ERL, while a high TRL corresponds to a high ERL. The central idea is that as a technology matures, its ethical implications become more concrete, increasing its ERL. At lower TRLs, high uncertainty and limited information result in a lower ERL, constraining the scope for meaningful ethical debate. Conversely, at higher TRLs, empirical data on the technology's effects becomes available, enabling detailed ethical scrutiny and supporting a broader range of methods.

The resulting framework, which systematically links Technology Readiness Levels (TRLs) to Ethics Readiness Levels (ERLs), can help guide assessments of the most appropriate ethical approach for a specific emerging technology. To do so, it is crucial to define the readiness levels at which a technology is considered "emerging." While there are more detailed definitions of "emerging technology" (e.g., Rotolo, Hicks, & Martin, 2015), the term is often used more generally in NEST-ethics to describe technologies that are still in development and not yet widely available, with their (ethical) impact expected to unfold in the future (Reijers et al., 2018). I will use this broader definition of "new and emerging technologies" and identify it with a specific range of TRLs.

As I discussed in Section 2**,** emerging technologies are typically understood as being in the early stages of development, characterised by significant uncertainty and novel possibilities. These characteristics are most pronounced at lower TRLs (1–3), where technologies remain in conceptual or early prototype phases, and mid TRLs (4–6), where viability is demonstrated, but full realisation is still underway. Thus, I propose that emerging technologies be defined as those at TRLs 1–6, where uncertainty is fundamental and broad in scope— acknowledging that the nature of this uncertainty shifts as a technology progresses from early-stage (TRL 1-3) to mid-stage (TRL 4-6). By the time technologies reach higher TRLs (7-9), they may be considered 'new' but have moved beyond the phase of emergence, as they are closer to full deployment and societal integration.

---

[5] While Urueña applies this concept to the qualities of anticipation frameworks, he also suggests that it can be applied to technologies themselves (2024: p.8).

At lower TRLs (1-3), uncertainty about whether, how, and when a technology will materialise dominates. This leads to a low ERL. As discussed in Section 2.4, this level of uncertainty diminishes the effectiveness of the outcomes-oriented ethical approach, which relies on anticipating specific impacts or consequences. In contrast, meaning-oriented ethics is less hindered by these unknowns, making it a more valid and practical approach at the earliest stages of technological development. In other words, the technology's critical-reflective affordances at this stage align with meaning-oriented ethics. This approach often implies an analytical focus on the broader level of the technology itself, instead of on specific designs and applications.

At mid TRLs (4-6), while the technology is still evolving and not yet fully realised, its technological viability has typically been demonstrated, and there is growing information about specific devices and potential applications. This raises the ERL and broadens the scope of ethical inquiry. As the technology starts to take more specific shape, it becomes both feasible and necessary to anticipate consequences. As a result, at this mid ERL methods within the outcomes-oriented approach gain relevance. This transition is accompanied by a shift in the level of analysis: from the broader focus on technology as a whole to a more specific consideration of artefacts and their applications. While the meaning-oriented approach remains valuable, outcomes-oriented methods increasingly complement it, reflecting the evolving affordances of the technology.

At higher TRLs (7–9), where the technology is practically deployed, its ERL is typically high. At this level, the relevance and effectiveness of meaning-oriented methods diminishes, while the need for outcomes-oriented methods intensifies. This is because, at this stage, real-world deployment leads to real-world impacts, making it crucial to anticipate and address ethical consequences in a concrete, outcomes-based manner. Moreover, the concreteness of the technology supports increasingly detailed ethical questions. Figure 2 summarises the ERL-framework.

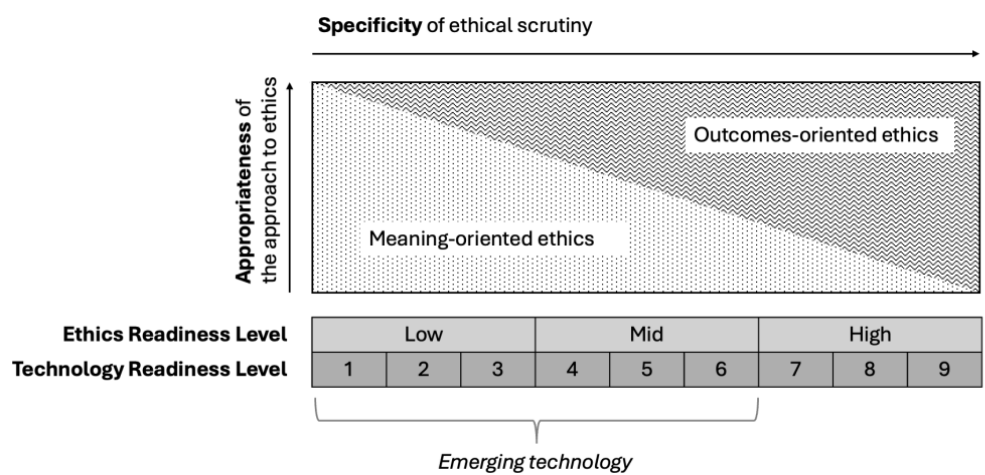

*Figure 2. Ethics Readiness Level (ERL) scale: The effectiveness and level of granularity of ethical approaches evolve along with technological maturation.*

This framework offers a broad guide for aligning the ethical approach to the specific stage of technological development. However, it serves as only a starting point. Further refinement of the ERL scale could enhance its ability to match specific ethical methods more precisely to specific stages of technology development. Additionally, as will be discussed in Section 4, other factors beyond a technology's ethics readiness may influence the ethical approach.

## 4. The ethics readiness of quantum technologies

In this section, I assess the ethics readiness of quantum technologies. As an emerging class of technologies, quantum technologies offer an interesting test case for the ERL scale, exploring its practical value and implications for the ethical study of emerging technologies.

Generally, quantum technologies refer to a family of technologies that leverage quantum effects[6] to create practical applications. These technologies are not entirely new; a first generation includes widely used devices like MRI scanners, lasers, and transistors. Recently, scientists and engineers have begun developing a second generation of quantum technologies, seeking to make further use of quantum phenomena by actively manipulating them in a more precise and controlled way. This promises new capabilities, such as enhanced sensing methods (*quantum sensing*), advanced computational power (*quantum computation*), and novel ways to communicate and secure information (*quantum communication*).

Although quantum ethics often discusses "quantum technology" as a single entity, it is important to distinguish between its subclasses, as they are at different stages of technological readiness. While not specifying exact levels, quantum computation and communication remain at the lower end of the technological readiness scale (TRLs 1-3). Despite significant breakthroughs in fundamental research, scalable prototypes are not yet available, and their viability not yet definitively confirmed. In contrast, quantum sensing has progressed further, falling between mid (TRLs 4-6) and high readiness levels (TRLs 7-9). This technology is proven feasible and close to ready for implementation in various applications.

Drawing on the idea that technology readiness levels (TRLs) correspond to specific ethics readiness levels (ERLs), the different TRLs of quantum technologies imply that they afford different ethical inquiries. The low technology readiness of quantum computing and quantum communication, indicates their generally low ethics readiness level. Given the suitability of the meaning-oriented approach to ethics at the earliest phases of development (Section 2.4 and 3.3), the ethical study of these technologies may thus generally be best approached by focusing on critical assessment of sociotechnical imaginaries and conceptual analysis. These technologies are still fundamentally 'in the making', making it difficult and speculative to assess their potential outcomes. However, the (emerging) ethical debate surrounding quantum computing and communication tends to predominantly focus on outcomes, emphasising potential (positive and negative) impacts. An important exception is the study by Inglesant et al. (2021), which explored current visions shaping quantum computing.

---

[6] Quantum technologies harness the principles of quantum mechanics, which is the theory that successfully describes how nature behaves at the scale of atoms and subatomic particles. At this scale, quantum phenomena occur, such as entanglement, superposition, and tunnelling. Quantum technology aims to leverage these phenomena for practical ends, which distinguishes them from classical technology.

Despite the early stage of quantum computer development, various stakeholders are already taking measures to mitigate specific risks, such as the potentially catastrophic impact of future large-scale quantum computers[7] on cybersecurity. These efforts include exploring and preparing for a transition to quantum-safe encryption—algorithms designed to resist attacks from quantum computers and safeguard digital data (NIST, 12 November 2024; TNO, CWI, & AIVD, 2023). At first glance, this anticipatory governance may seem premature given the nascent state of quantum computing. However, the potentially severe consequences of a quantum cybersecurity breach may provide a rationale for anticipating specific outcomes and, in turn, for considering an outcomes-oriented ethical approach, even amid profound uncertainties (Price, 2024). Still, such an exception depends on well-supported, realistic expectations about the technology's trajectory and its potential consequences (Nordmann & Rip, 2009). Further research should explore whether and under what conditions outcomes-oriented ethics may be justifiable at a low TRL.

Quantum sensing, on the other hand, occupies a higher TRL and thus a higher ERL. With its technological viability demonstrated, it affords more detailed and specific ethical inquiries of impacts. Outcomes-oriented ethics becomes applicable at this stage and gains relevance, as demonstrated technological feasibility implies the urgency of anticipating potential consequences. However, based on the output of academic literature, the current ethical debate about quantum technologies seems to gravitate toward quantum computing and communication, while relatively neglecting sensing. This may reflect a tendency to prioritise compelling yet speculative scenarios related to quantum computing and communication, at the expense of addressing the ethical implications of the more near-term applications of quantum sensing.

The difference in ethical focus suggested for different types of quantum technologies illustrates how ERLs help identify when and how different ethical approaches become relevant as a technology matures. By aligning the ethical approach with the stage of technological development, we can ensure that ethical discussions remain relevant and appropriately grounded at every phase of a technology's evolution.

## 5. Conclusion

Over the past decades, numerous methods have emerged for the ethical study of emerging technologies, each grappling in their distinct ways with the inherent uncertainties of innovation. As technologies progress, however, the nature of these uncertainties evolves, suggesting that ethical approaches should adapt to align with specific stages of technological development. Yet, until now, an overarching framework to evaluate the suitability of different ethical methods based on a technology's stage of development, is lacking.

In this paper, I aimed to address this gap. Considering the vast number of methods that have been developed for the ethics of emerging technology, I distinguished between two main approaches: outcomes-oriented ethics—typically focused on the ethical assessment of consequences—and the meaning-oriented ethics—typically focused on the critical reflection on motives or intentions driving technological projects. Since these approaches each deal in their own way with the uncertainty that characterise early-stage technologies, I suggested that their appropriateness and relevance depend on the availability of information, and thus on the specific stage of technological development.

---

[7] A cryptographically relevant quantum computer (CRQC) is a quantum computer with a sufficient number of reliable qubits and error-correcting capabilities to efficiently break widely used encryption methods, such as RSA or ECC, by solving the underlying mathematical problems (Mosca & Piani, 2024).

To substantiate this idea, I introduced the concept of *ethics readiness*: the readiness of a technology to undergo detailed ethical scrutiny. I argued that the ethics readiness of a technology evolves along with technological maturation, determining both the effectiveness of outcomes- and meaning-oriented ethics, as well as the supported level of granularity or specificity of ethical analysis. To make ethics readiness more concrete, I defined three Ethics Readiness Levels (ERLs), in alignment with the Technology Readiness Levels (TRLs) scale. The resulting framework indicates that early-stage technologies (TRL 1-3) typically exhibit a low ERL, reflecting the limited capacity of an immature technology to support detailed ethical analysis. At this stage, when uncertainty is fundamental and broad, the meaning-oriented approach to ethics is best suited. When a technology progresses to higher TRLs (4-9) and more information becomes available, it supports and even calls for an outcomes-oriented approach to ethics, allowing for increasingly specific ethical scrutiny.

The ERL-framework thus depicts a shift in the appropriateness of outcomes- and meaning-oriented approaches to ethics as a technology progresses and takes shape more concretely. The iterative application of the framework to the nascent field of quantum technologies offered a first exploration of its implications. This exercise demonstrated that according to this framework, different types of quantum technologies call for different ethical approaches based on their specific TRLs.

By linking technological readiness with ethical readiness, the ERL framework ensures that ethical reflection evolves in tandem with technological advancement. It offers a dynamic, pragmatic approach to upstream ethics, aligning ethical reflection with the affordances of technology's maturity. In doing so, it reframes the tension between engaging too early and too late—navigating between the Scylla of premature speculation and the Charybdis of missed ethical opportunities. In that way, ERLs lay the groundwork for a more responsive and impactful upstream ethics.